\documentclass[achemso,superscriptaddress, twocolumn]{revtex4-2} 
\usepackage{mathtools}
\usepackage{amsmath}
\usepackage{multirow}
\usepackage[english]{babel}
\usepackage{amsmath,lipsum}
\usepackage{etoolbox}
\appto{\normalsize}{\zerodisplayskips}
\appto{\small}{\zerodisplayskips}
\appto{\footnotesize}{\zerodisplayskips}
\usepackage{textcomp}
\usepackage{gensymb}
\newcommand{\zerodisplayskips}{%
  \setlength{\abovedisplayskip}{5pt}%
  \setlength{\belowdisplayskip}{5pt}%
  \setlength{\abovedisplayshortskip}{5pt}%
  \setlength{\belowdisplayshortskip}{5pt} }
\appto{\normalsize}{\zerodisplayskips}
\appto{\small}{\zerodisplayskips}
\appto{\footnotesize}{\zerodisplayskips}
\usepackage{graphicx}
\usepackage{color}
\usepackage{enumitem}
\usepackage[dvipsnames]{xcolor}
\usepackage[normalem]{ulem}

\usepackage{float}
\usepackage{booktabs}
\usepackage{tabularx}
\usepackage{subcaption}
\usepackage{hyperref}
\renewcommand\thesubsection{\arabic{subsection}}
\renewcommand\thesubsubsection{\arabic{subsubsection}}
\makeatletter
    \def\@seccntformat#1{\@ifundefined{#1@cntformat}%
       {\csname the#1\endcsname\space}
       {\csname #1@cntformat\endcsname}}
    \def\subsection@cntformat{\thesection.\thesubsection\space} 
    \def\subsubsection@cntformat{\thesection.\thesubsection.\thesubsubsection\space}
\makeatother

\usepackage{fancyhdr}
\usepackage{blindtext}
\DeclareUnicodeCharacter{0301}{\'{e}} 
\DeclareUnicodeCharacter{2009}{\,} 
\DeclareUnicodeCharacter{2212}{-}
\DeclareUnicodeCharacter{2500}{\mbox{\vrule height2.2ptdepth-1.8ptwidth.5em}}

\begin{document}

\title{Simulating Crystallization in a Colloidal System Using State Predictive Information Bottleneck based Enhanced Sampling}

\author{Vanessa J. Meraz}
\affiliation{Institute for Physical Science and Technology, University of Maryland, College Park 20742, USA.}

\author{Ziyue Zou}
\affiliation{Department of Chemistry and Biochemistry, University of Maryland, College Park 20742, USA.}

\author{Pratyush Tiwary\thanks{ptiwary@umd.edu}}
\email{ptiwary@umd.edu}
\affiliation{Institute for Physical Science and Technology, University of Maryland, College Park 20742, USA.}
\affiliation{Department of Chemistry and Biochemistry, University of Maryland, College Park 20742, USA.}
\affiliation{University of Maryland Institute for Health Computing, Rockville 20852, USA.}

\date{\today}

\begin{abstract}
\section*{Abstract}
\label{sec:abstract}
We investigate crystal nucleation in supersaturated colloid suspensions using enhanced molecular dynamics simulations augmented with machine learning techniques. The simulations reveal that crystallization in the model colloidal system studied here, with particles interacting through a repulsive screened Coulomb Yukawa potential, proceeds from vapor to dense liquid droplet to crystalline phases across multiple high barriers. Employing a one-dimensional reaction coordinate derived from the State Predictive Information Bottleneck framework, our simulations capture back-and-forth phase transitions across multiple barriers effectively in biased metadynamics simulations. We obtain relative free energy differences between different phases and also quantify the roles of different molecular level features in driving the phase changes. 
\end{abstract}

\maketitle
\thispagestyle{fancy}

\section{Introduction}
\label{sec:introduction}

Fundamental to physical chemistry, nucleation is critical in the study of kinetics as the key initiation event of crystallization. Effectively controlling the nucleation process of molecules or particles within a solvent is of great importance amongst different scientific disciplines \cite{Sosso2016microscopic, Fersht1997, Aizenberg2004, Baumgartner2013}. In all cases, a cluster forms within the solvent and when it is sufficiently large enough, the free energy barrier is overcome by the gain of emerging crystalline order. While ubiquitous in nature and well observed macroscopically, the mechanism behind this event is not so well understood. Even at the start of the process, the initial crystallization greatly influences the polymorph produced \cite{Zou2021, Zou2023}. Generally, the basis for studying nucleation is built upon classical nucleation theory (CNT), a one-step theoretical model, which considers only one variable necessary to capture the complex dynamics resulting in phase transformations: the size of the largest cluster, assumed to be spherical \cite{Kal2016, Sosso2016}.

Though simple and effective, CNT has been continuously improved upon, the most prominent and extensively studied modification of which has been the two-step model \cite{Kal2016}. Through the simplistic view of CNT, there is a single free energy barrier that needs to be overcome for crystallization to be energetically favorable. The modified two-step mechanism accounts for metastable states between the starting supersaturated solution and the emerging crystal. Hence, there is more than one free energy barrier whose cost needs to be offset by crystal growth. This idea of multistep crystallization is motivated by the importance of density fluctuations when clusters are being formed. For example, the colloidal system studied in this work starts as a vapor phase which can then grow into a dense liquid droplet (DLD) if local density fluctuations promote the clustering of particles which then may or may not favor the ordering necessary for crystal emergence. While we can generalize the steps for a system to nucleate, the path for crystallization itself is not unique. As such, molecular dynamics (MD) simulations could be particularly useful in directly observing crystallization \cite{Finney2022}. However, nucleation is a rare event whose timescale varies amongst systems (within seconds to minutes) and is typically longer than what is accessible and achievable by MD. 

In this work, we study crystallization in supersaturated colloid suspensions \cite{Cacciuto2004}. The particles in our system interacted through a repulsive screened Coulomb Yukawa potential \cite{yukawa}, which are commonly used to describe charge-stabilized colloidal suspensions \cite{PhysRevLett.57.2694}. Due to their larger size, colloids make for a challenging model system in the study of nucleation. In some sense, their slow movement allows for an easier look into dynamics that may not be all apparent in atomistic systems. However the interaction between colloidal particles themselves presents unique challenges. Specifically, in a stable suspension, there is weak interaction between colloidal particles lending to the difficulty of spontaneous aggregation into crystals. 

The crux of the problem in understanding nucleation lies in the atomistic detail necessary to glean information as to what promotes or inhibits it, a formidable feat even for MD with state-of-the-art instrumentation. For this reason, we turn to enhanced MD algorithms to reproduce the crystallization process. Unsurprisingly, there have been several efforts in attempt to sample a larger portion of configuration space within the allotted simulation time \cite{shams2024}. One enhanced sampling approach involves learning an approximation to the true, \textit{a priori} unknown, reaction coordinate (RC) that can represent the progress along a generic reaction pathway. Generally, such an RC is constructed as a function of collective variables (CVs), or order parameters (OPs) \cite{matteo2023, Neha2022, PrezHernndez2013, Trout2011orientationop}, which are functions of particles’ positional coordinates. For enhanced sampling of rare events, one then constructs biasing potential or forces as a function of the RC, thereby speeding up slow processes. An important caveat to note about the use of OPs used to build the RC is their requirement to be continuous and differentiable.

In this work, we bias the approximate RC through the well-tempered metadynamics method \cite{Laio2002, Laio2008, PRL2008WTMetaD, matteo2016cn, bussi2020using, shams2024}. The algorithm artificially fills free energy wells with a bias potential as a function of the approximate RC enabling an enhanced exploration of the landscape. We use the State Predictive Information Bottleneck (SPIB) framework to construct the RC \cite{Wang2021}. In the most broad sense and in methods like SPIB, the RC is constructed as a linear or nonlinear combination of OPs \cite{PrezHernndez2013, sgoop,ma2005automatic, best2005reaction, baron2006obtaining}. Using OPs that are computationally cost-effective to calculate and differentiate, as decided by their ranking detailed in Ref. \cite{matteo2023}, a linear RC is constructed to best enhance crystallization. We find that such a one-dimensional RC is sufficient to observe back and forth phase transitions in metadynamics simulations. Through this, we observe different metastable states relevant to crystallization in the colloidal system. Starting from the vapor phase, we find that the system first goes to a dense liquid droplet phase, followed by a crystal phase. We observe multiple back-and-forth transitions between these phases. Additionally, oftentimes the system continues to a more dense crystal phase, at which point it gets trapped there. This behavior reflects both the strengths and limitations of our SPIB based RC. By reweighting the well-tempered metadynamics simulations, we also calculate the relative free energy differences between these different metastable states. 

While machine learning based reaction coordinate methods\cite{shams2024} have been recently applied to a large number of biophysical systems, their application to nucleation in colloidal systems is rather limited. Our work shows that these methods can also be directly applied to the study of nucleation in model colloids. We thus believe the protocol demonstrated in this work will be useful for gathering thermodynamic insights into generic colloidal systems.

\section{Methods}
\label{sec:method}
The development of enhanced sampling methods has been of great use in the atomistic simulation of crystal nucleation which would otherwise be improbable to observe even with the best available supercomputers. In this work, we are interested in studying nucleation in a system of supersaturated colloid suspensions using a Yukawa potential \cite{yukawa2}.  To simulate nucleation and obtain accurate associated thermodynamic observables, here we use the enhanced sampling method well-tempered metadynamics \cite{PRL2008WTMetaD} biasing an approximate RC learnt from an SPIB model. 

\subsection{Nucleation Order Parameters} 
\label{sec:op}
SPIB constructs an approximate reaction coordinate (RC) as a function of input features or order parameters (OPs). Through there isn't a clear-cut definition for which OPs to use for studying nucleation, there have been extensive studies on the relative importance of different possible OPs \cite{matteo2023, Trout2011orientationop, Parrinello2021cv}. In this text, we focus on OPs studied in ref. \cite{matteo2023}, and in particular those of which we find to be cost effective in terms of simulation time used to compute these OPs and forces acting on them. They are respectively classified into two categories: average properties of all particles (energy, coordination number, Steinhardt bond orders) and total number of particles according to local geometry (particles in a condensed or solid-like phase). They are listed in Table \ref{tab:op} for reference and are defined as follows: 

\begin{table}
   \centering
   \caption{Order parameters (OPs) input to SPIB}
   \begin{tabular}{c|c}
   \hline
   \hline
         Label & OP \\
         \hline
         cn.mean & Mean particle first-sphere coordination number \\
         ncl &  Number of particles in a condensed phase (CN $>$ 3) \\
         ncs &  Number of particles in a solid-like phase (CN $>$ 6) \\ 
         q4.mean &  Mean Q4 Steinhardt bond order \\
         q6.mean &  Mean Q6 Steinhardt bond order \\
         ene &  Potential energy\\
         \hline
         \hline
    \end{tabular}
    \label{tab:op}
\end{table}

\begin{enumerate}[leftmargin=*]
\item Coordination Number: \textit{cn.mean} \newline \\
Eq.\ref{eq: cn_avg} is used to compute the mean first-sphere coordination number, $cn.mean$, where $r_{ij}$ is the distance between particles and $r_{max}$ is the distance within the first-coordination sphere. It follows from the general definition of a coordination number for a particle $i$ as given by Eq. \ref{eq: coord_sphere} where $r_o$ is a cut-off radial distance.
\begin{equation}
    cn.mean = \frac{1}{N}\sum_{i,j\neq i}^N \frac{f(r_{ij}) - f(r_{max})}{1 - f(r_{max})}
    \label{eq: cn_avg}
\end{equation}

\begin{equation}
    f(r_{ij}) = \sum_{j} \frac{1 - (r_{ij}/r_o)^6}{1 - (r_{ij}/r_o)^{12}}
    \label{eq: coord_sphere}
\end{equation}
Here the summation variable $j$ enumerates all other particles. A general descriptor of a particles’ local environment, the coordination number is a simple number OP. It fails to capture the symmetry of a crystal structure and thus is not able to distinguish emerging phases making it inadequate as a stand-alone OP. 

\item Number of particles in a condensed or solid-like phase: \textit{ncl, ncs}\newline \\
Calculating the number of particles in a phase is similar to the description of the coordination number $cn$. The $cn$ is calculated over all particles, for which there is a distribution, and a filter, 1− f(r = x), is then applied to classify the phase as condensed or solid-like. Bounds x=3 and x=7 were used to find \textit{ncl} and \textit{ncs} respectively. The term f(x=r) is the same as seen in Eq. \ref{eq: coord_sphere}.

\item Steinhardt bond order parameters: \textit{q4.mean, q6.mean} \newline \\
To quantify the symmetries in a given cluster made of different particles, Steinhardt bond order parameters are widely used \cite{steinhardt}. These OPs are functions of spherical harmonics and describe the ordering of the local environment surrounding a given particle. Here, we set the \(l^{th}\) order of the spherical harmonics $\mathbf{Y_{lm}}$ to 4 and 6 as they result in interesting characteristics of crystal structures \cite{Lechner2008}. Accordingly, the \(m^{th}\) order has ranges \{-4,...,+4\} and \{-6,...,+6\}  with respect to the \(l^{th}\) order. This results in 9 or 13 spherical harmonic functions to be evaluated as seen in Eq. \ref{eq:sh}, where $i,j$ are the same as in Eq. \ref{eq: cn_avg}:
\begin{equation}
    q_{lm} = \frac{\sum_j r(r_{ij}) \bf{Y}_{lm}(\bf{r}_{ij})}{\sum_j r(r_{ij})}\label{eq:sh}
\end{equation}

\item Potential energy: \textit{ene}\newline \\
Each time step of a simulation requires the calculation of potential energy as a function of the force field chosen. As we intend to study phase transformations, the potential energy of a system is a good indicator one has taken place. As seen in Fig. \ref{fig: ene_ts_crystal}, the increase in order of a system (from vapor-to-liquid or liquid-to-crystal) results in a significant potential energy difference. In the system studied in this work, a colloid/Yukawa \cite{yukawa2} potential was used. 

\end{enumerate}

\subsection{State-Predictive  Information  Bottleneck  (SPIB)}
\label{sec:spib}
At the heart of understanding nucleation, and countless other phenomena in physical chemistry, sits the need to project very high-dimensional data onto a much lower dimensional manifold. To effectively address this task, it's crucial to have a concise and informed representation capable of distinguishing between different states within a system, as well as capturing a correct state-to-state transition probability density. As previously discussed, a multitude of variables are necessary for accurately describing transitions, hence a combination of OPs is employed to craft such a representation. Within this framework, we employ the State Predictive Information Bottleneck (SPIB), functioning akin to a variational autoencoder (VAE).

SPIB serves to preserve pertinent dynamics between metastable states by compressing a high-dimensional Molecular Dynamics (MD) trajectory into a low-dimensional representation of latent space which can be shown to approximate certain critical requirements of the true RC \cite{Wang2022}. This representation extracts the essential information from an MD trajectory, minimizing redundancy while maximizing its predictive power regarding future states following a specified time delay, denoted as $\Delta t$. 

For the RC to be able to learn the number of metastable states and differentiate between them, a variational mixture of posteriors prior (VampPrior)\cite{vampprior} is used to construct a multimodal probability distribution prior for SPIB. The prior is approximated as a weighted combination of, in our case, OPs with initialized pseudo-inputs to be updated on-the-fly during training. For a particular choice of time delay $\Delta t$, iterations of SPIB will converge to an $N$ number of metastable states whose population probability is reflected in the updated pseudo-inputs. Generally speaking, the number of converged metastable states decreases as the time-delay $\Delta t$ is increased. See Ref. \cite{Wang2021} for further description of SPIB.

The choice of encoder, linear or not, is left to the application. While a non-linear encoder can be more analytically expressive, the need to interpret results from biased metadynamics simulations restricts our use to a linear encoder. Furthermore, for poor quality training data a non-linear encoder can also be prone to overfitting and less capable of extrapolation outside training domain, which is often critical and especially so in biased simulations.

\subsection{Enhanced Sampling with Metadynamics}
\label{sec:metad}
Since nucleation is a rare event which is difficult to simulate directly in unbiased MD, one needs to use enhanced sampling methods to simulate them. Here we use well-tempered metadynamics. Metadynamics is a technique used to explore a broader portion of the configuration space by applying a history-dependent bias in the free energy landscape, described as a function of one to two carefully chosen biasing variables. In the well-tempered variation of metadynamics \cite{PRL2008WTMetaD}, the height of the Gaussian added at any given point in biasing variable space is gradually reduced during the simulation. Gaussian biasing potentials are periodically introduced as function of these biasing variables, or the approximated RC, to discourage the system from getting stuck in local free energy minima.  The net product of a converged metadynamics simulation is the free energy as function of the biased or any other variables through a reweighting procedure \cite{Tiwary2014}. While metadynamics is mathematically proven to converge for any biasing variable \cite{Dama2014}, in practice the speed of convergence is improved if the variables being biased capture all relevant slow degrees of freedom, i.e. approximate the true reaction coordinate \cite{bussi2020using}. Here we achieve this through the use of SPIB. 
 
Specifically, we use SPIB to generate a one-dimensional RC as a linear combination of input OPs provided in Table \ref{tab:op} which requires initial trajectories that have seen at least one phase transition. This can be obtained by either biasing a trial RC or by launching a very large number of unbiased runs and selecting one where nucleation happens. Here we do so by running 1000 independent unbiased simulations, 21 of which show nucleation, and hence are used to create this RC. Having sampled this space, the metastable states of the system will be able to be discretized with some degree of accuracy.

In Table \ref{tab:metaD}, we report the parameters used for well-tempered metadynamics simulations. These are (i) the initial height of the deposited Gaussian $\omega$ in units of $k_BT^*$, (ii) bias factor $\gamma$ (unitless), (iii) width of the Gaussian $\sigma$ in same units as the SPIB RC being biased; and the reduced temperature $T^*$ defined as $k_BT$ / $\epsilon$ . The Gaussian width $\sigma$ was kept at 5 times from the standard deviation of the RC measured in a highly paced biased run to simulate an unbiased one which lasted 3.25 x $10^5$ steps with timestep $\Delta$t = 0.005.

\begin{table}[h]
   \centering
   \caption{Parameters for metadynamics (Definitions and units are provided in Sec. \ref{sec:metad})}
   \begin{tabular*}{0.8\columnwidth}{@{\extracolsep{\fill}} *{4}{c} }
   \toprule
   $\omega$ & $\gamma$ & $\sigma$ & $T^*$\\
   \midrule
   2.0 & 50 & 1.3781 & 2\\
   \bottomrule
   \end{tabular*}
   \label{tab:metaD}
\end{table}

\section{Results}
\label{sec:results}

\subsection{Observations from initial Molecular Dynamics Simulations}
\label{sec:sim_init} 
The colloidal simulations performed in this work were done using LAMMPS \cite{lammps} patched with PLUMED \cite{plumed2019nature, plumed2}, following the detailed set-up reported in Ref. \cite{matteo2023}. 

\begin{figure}[ht] 
  \centering
  \includegraphics[width=.45\textwidth]{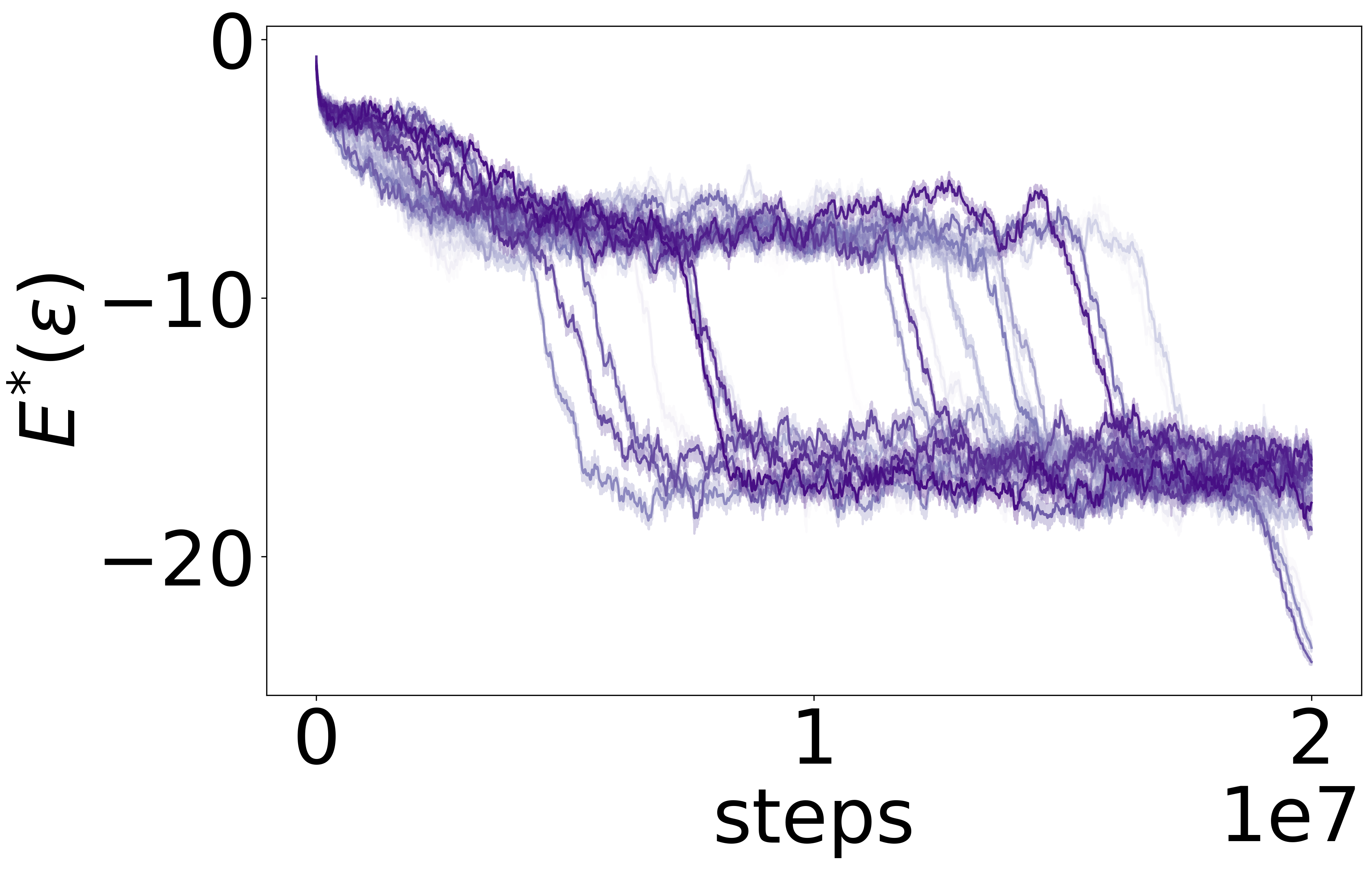}
  \caption{Mean potential energy in reduced units, $E^*$, per particle as a function of time measured in MD integration timesteps for all 21 crystallized trajectories illustrated in different shades of purple.}
  \label{fig: ene_ts_crystal}
\end{figure}

We began by launching 1000 independent unbiased MD simulations starting from the vapor phase to gather instances where crystallization occurred. As in Ref. \cite{matteo2023}, this was characterized through the polyhedral template matching (PTM \cite{Larsen2016}) tool in LAMMPS. Following that same protocol, we impose the condition that the crystal structure formed during the simulation must consist of a minimum of 188 particles. This condition adheres to the threshold value for potential energy per particle E* = -10 $\epsilon$ seen in Fig.\ref{fig: ene_ts_crystal}. Of the initial 1000 simulations, only 21 trajectories formed a crystalline structure. This is similar to the crystallization rate around 1\% observed in Ref. \cite{matteo2023}, in tune with the rare event nature of colloid nucleation. 

Furthermore, none of these select successful unbiased trajectories in our work or in Ref.  \cite{matteo2023}, once crystallized, ever revisited the starting vapor phase. This illustrates that unbiased simulations such as these are not capable of providing equilibrium free energy profile for the colloidal system.

\begin{figure}[htbp]
    \centering
    \begin{subfigure}[b]{0.4\textwidth}
        \centering
        \includegraphics[width=\textwidth]{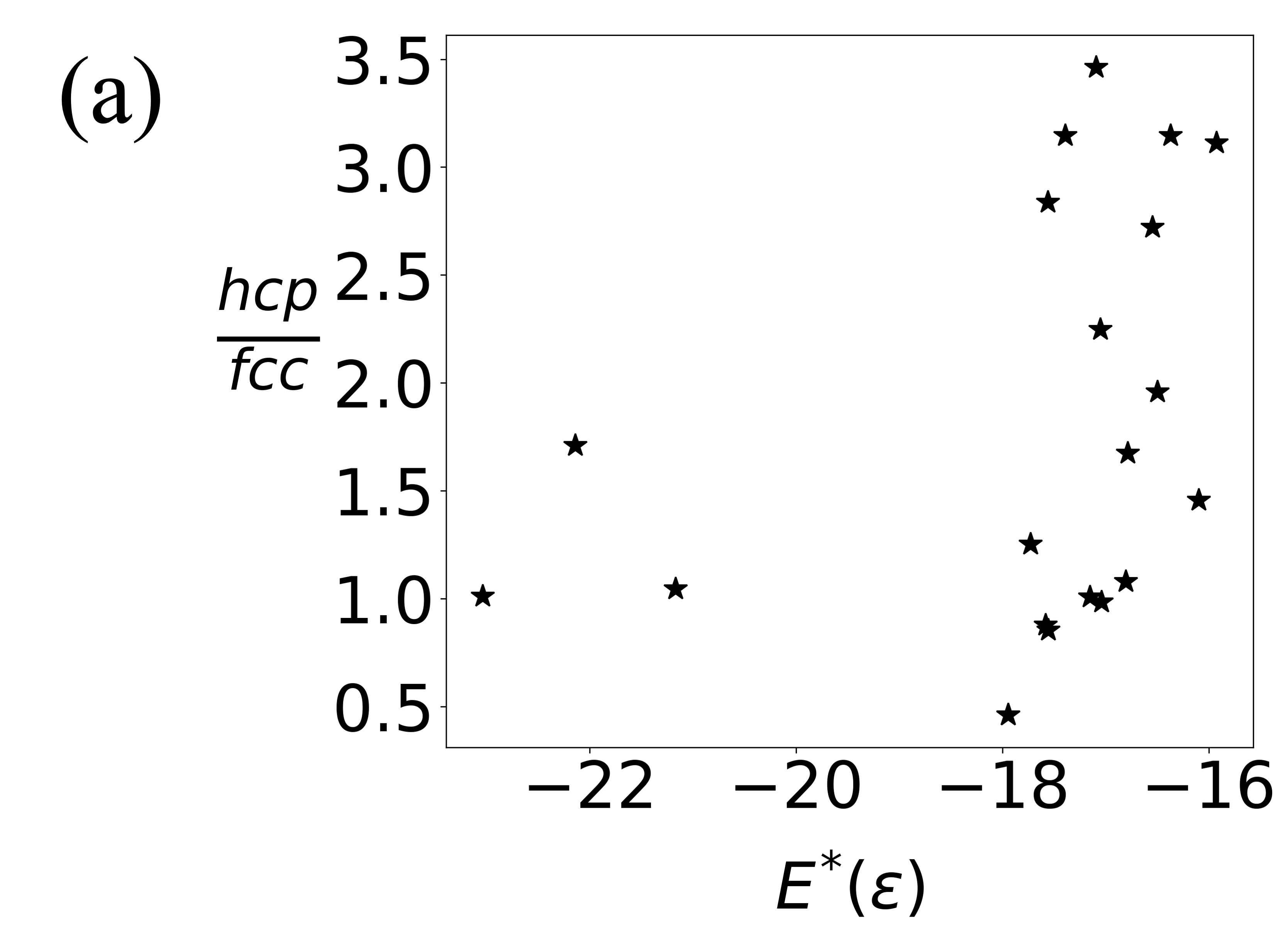}\label{fig: orig_ratio}
    \end{subfigure}
    \quad
    \begin{subfigure}[b]{0.4\textwidth}
        \centering
        \includegraphics[width=\textwidth]{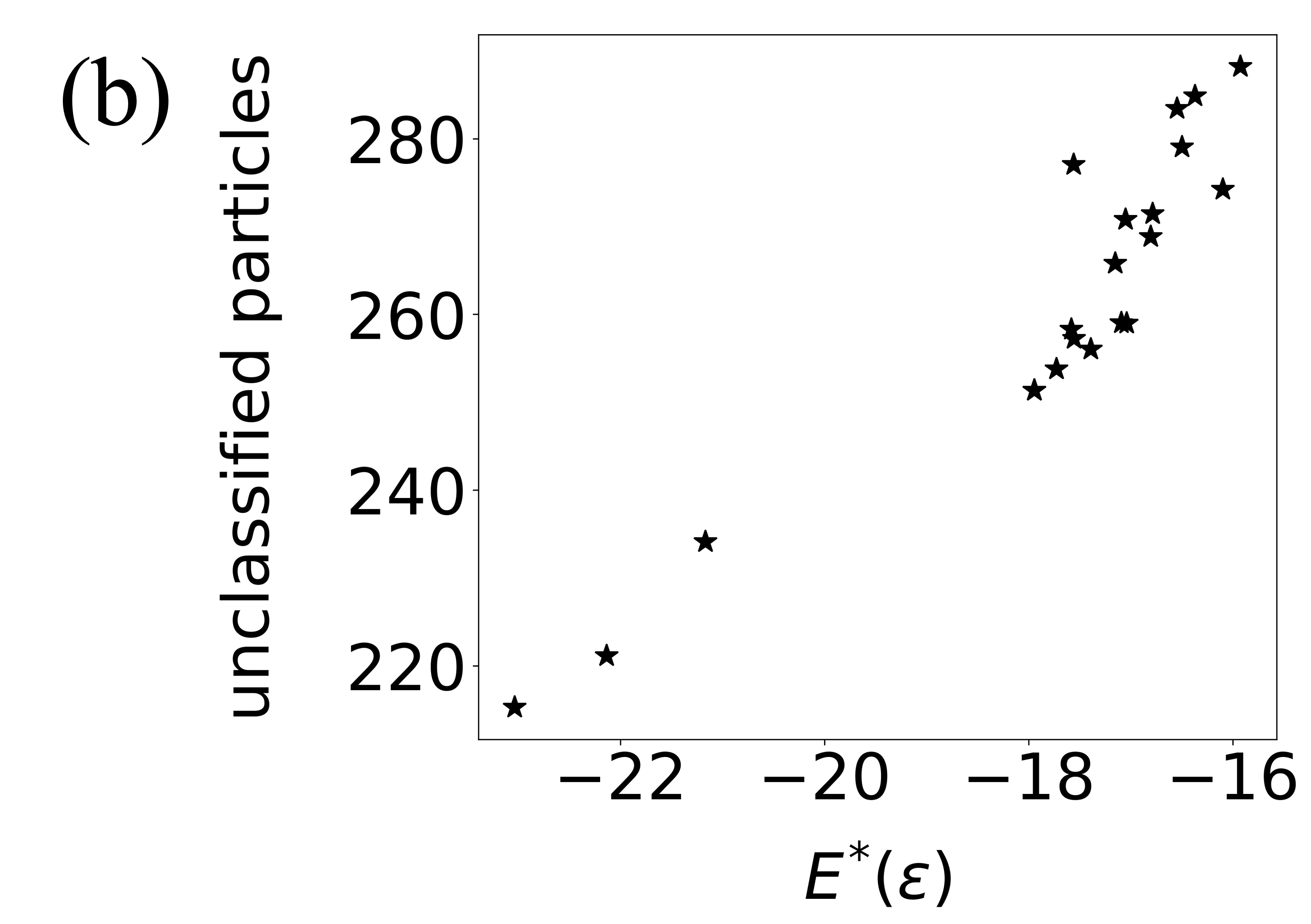}\label{fig:unclassified}
    \end{subfigure}
    \caption{(a) Ratio of hcp/fcc crystals plotted against $E^*(\epsilon)$ which is the average of potential energy measured over the last 5,000 frames.  \\ (b) Number of particles not classified as fcc or hcp out of total 388 particles plotted against $E^*(\epsilon)$.}
    \label{fig:ratios}
\end{figure}

In Fig. \ref{fig: ene_ts_crystal}, we plot the mean potential per particle as a function of MD integration timesteps for the 21 unbiased MD trajectories that showed crystallization. As was also reported in Ref. \cite{matteo2023}, there are clear features representative of phase change. Of the 21 trajectories, 3 are of particular interest due to their formation of a densely packed structure, with minimal particles excluded. In Fig. \ref{fig: ene_ts_crystal}, these three instances can be seen where the mean potential energy goes below E* = -20 $\epsilon$. To understand this phenomenon better,  we plot the ratio of particles classified as hexagonal close-packed (hcp) versus face-centered cubic (fcc) in Fig. \ref{fig:ratios}(a).  The 3 lowest energies recorded in Fig. \ref{fig: ene_ts_crystal} correspond to the 3 structures seen as outliers in the bottom left of Fig. \ref{fig:ratios}(a). They have the same relative hcp/fcc ratio but markedly lower potential energy. This is further seen in Fig. 2b where the number of particles not classified as fcc or hcp are plotted against the reduced mean potential energy and the three points of interest seemingly behave as outliers.    

\begin{figure}[h!]
    \includegraphics[width=.8\linewidth]{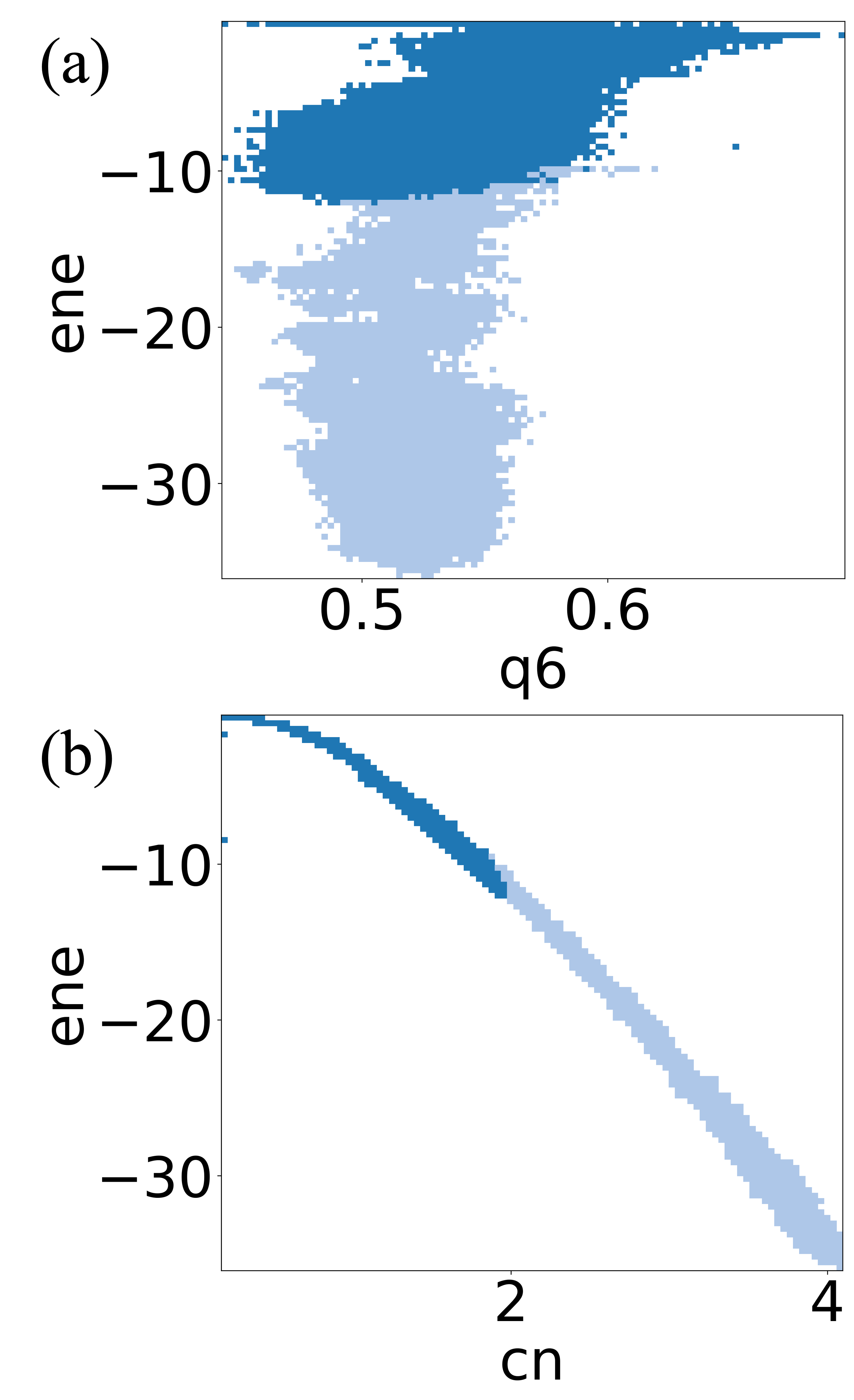}
    \caption{Projection of SPIB RC onto different OPs. In (a) we plot the average potential energy ene versus the Steinhardt bond order parameter q6. In (b) we plot the average potential energy versus the coordination number cn. In both (a) and (b), the different shades of blue indicate the two states SPIB converged to given a time lag of $\Delta t$ = 10,000.}
    \label{fig: spib_ene}
\end{figure}

To construct the RC for biasing in metadynamics, the order parameter outputs from these crystallized trajectories are given as training input data into SPIB. The learned RC is constructed as a one-dimensional linear encoder as a function of OPs described in Sec. \ref{sec:op}. For simplicity, we choose a time lag long enough ($\Delta t$ = 10,000 MD integration timesteps) for states to be discretized between crystal and non-crystal basins, as seen with distinct colors in Fig. \ref{fig: spib_ene}. The projection of the RC along different order parameters seen in the figure also illustrates the rationale behind the previously defined threshold for crystallization (around E* = -10 $\epsilon$). 

\subsection{Observations from Metadynamics biasing SPIB}
\label{sec:sim}
\begin{figure}[h!] 
  \centering
  \includegraphics[width=.4\textwidth]{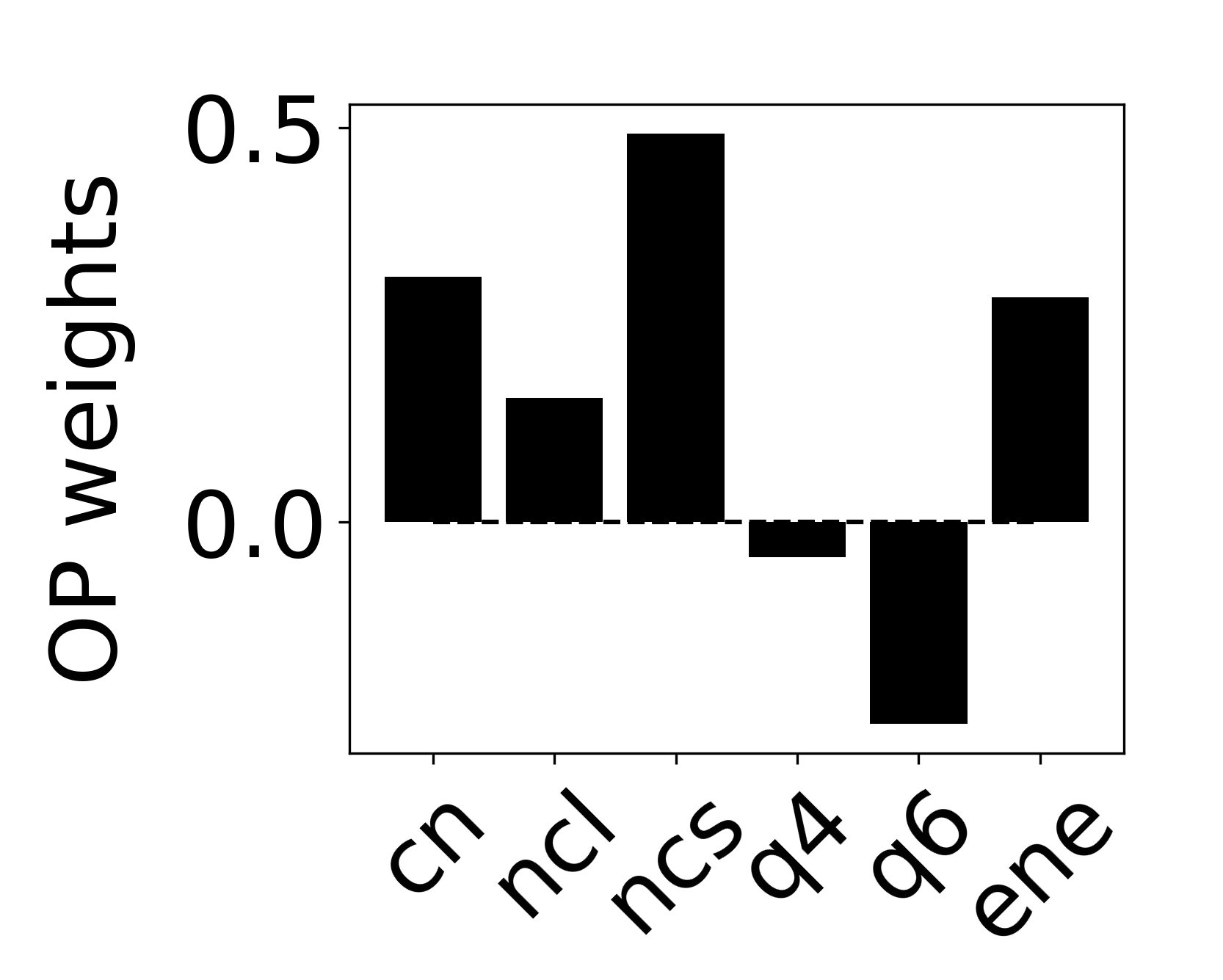}
  \caption{Order parameter (OP) weights from RC obtained through SPIB.}
  \label{fig: RC weights}
\end{figure}

As constructed by the one-dimensional linear SPIB encoder, the RC is expressed as a combination of the OP weights seen in Fig. \ref{fig: RC weights}. From the figure, we note the high contribution of the OP capturing number of particles in a solid-like phase (ncs) followed by the mean coordination number (cn), average potential energy, and Steinhardt bond order parameter q6. Conversely, the Steinhardt bond order parameter q4  contributes the least.

\begin{figure}[h!]
    \centering
    \vspace{0.5cm}
    \includegraphics[width=\linewidth]{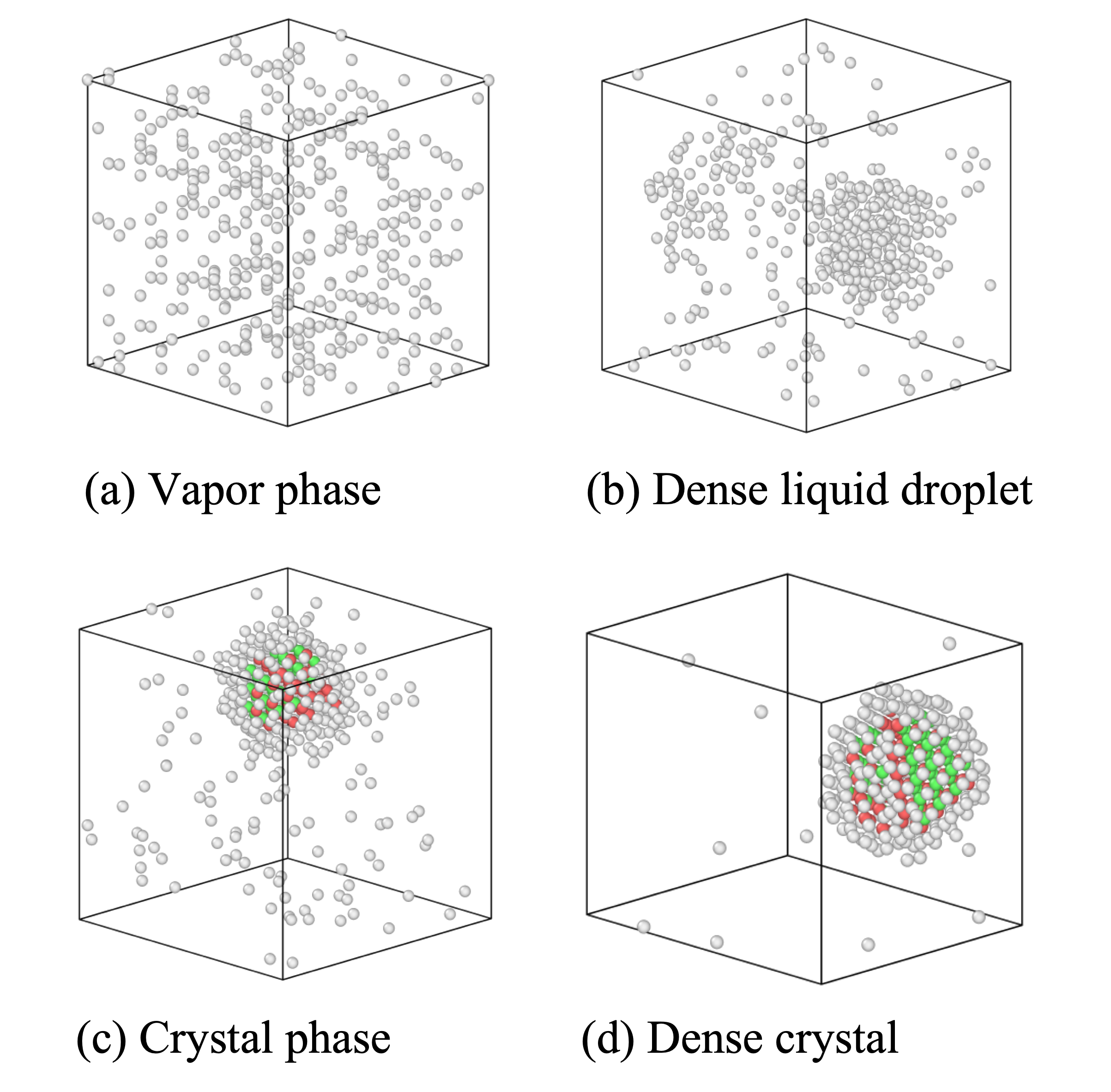}
    \caption{Phases seen during MD simulation as rendered by Ovito. Within sub-figures (c) and (d), particles in red are in the hcp phase and particles in green are in the fcc phase.}
    \label{fig:phases}
\end{figure}

We performed 10 independent well-tempered metadynamics simulations biasing along the SPIB based RC, which we label $z$. In contrast to the original $\sim$2\% rate of crystallization observed in the initial unbiased simulations, all metadynamics runs indicated crystal formation as well as back-and-forth movement between some phases. Fig. \ref{fig:phases} shows the 4 different phases of the system observed during our metadynamics simulations. The categorization of these 4 phases as a function of the SPIB RC, $z$, is as follows: vapor, for $z <$ 20; dense liquid droplet, for 20 $< z <$ 40; crystal, for 50 $< z <$ 70,  and a dense crystal phase for $z > 90$.

\begin{figure}[h!]
    \centering
    \vspace{0.5cm}
    \includegraphics[width=.9\linewidth]{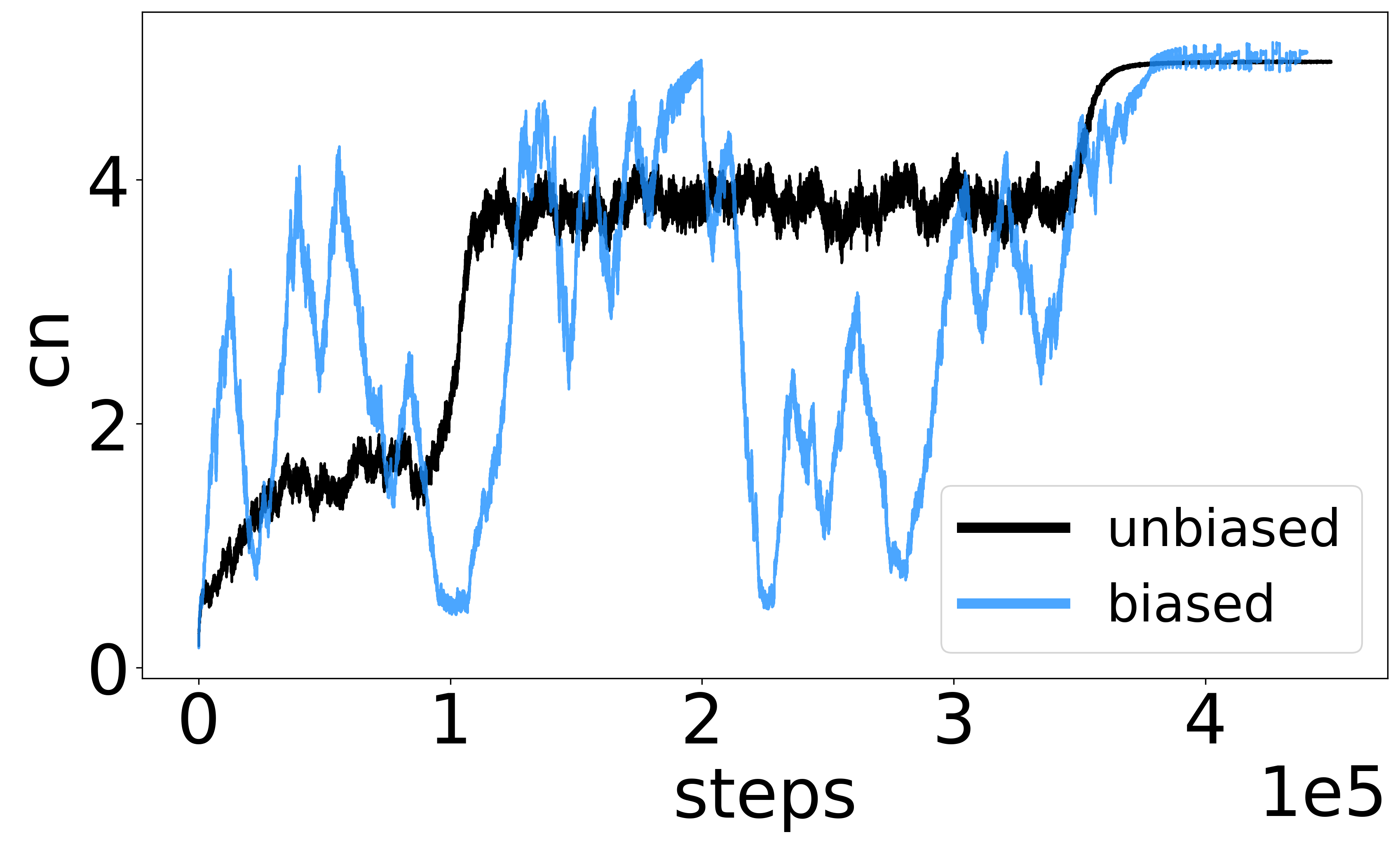}
    \caption{Coordination number cn as a function of simulation time (measured in MD timesteps) for typical biased and one of the successful unbiased trajectories, shown in colors blue and black respectively.}
    \label{fig:cn_metad}
\end{figure}

In Fig. \ref{fig:cn_metad} we plot the coordination number, cn, as a function of simulation time for a typical metadynamics trajectory (see SI for other nine metadynamics trajectories showing qualitatively similar behavior). This figure shows a clear back and forth movement between 3 of the 4 phases mentioned in the previous paragraph. In other words, our trajectories move back-and-forth between the vapor, dense liquid and crystal phases. However, for 8 of the 10 trajectories, the system reaches a dense crystal phase with $z >90$. 

If this phase is reached, our simulations do not revert back. This reflects that the RC learnt through SPIB captured the physics relevant to moving between the first three phases including the crystal, but not for movement back from the dense crystal. One possible explanation for this is the under-representation of trajectories visiting the dense crystal phase in our training data, where only 3 out of 21 trajectories visited this phase. In Section \ref{sec: fep} we elaborate qualitatively on the basis of the free energy profile.

\newpage
\subsection{Free Energy Profile}
\label{sec: fep}
\subsubsection{Initial Molecular Dynamics Simulations}

\begin{figure}[h!]
    \centering
    \includegraphics[width=.7\linewidth]{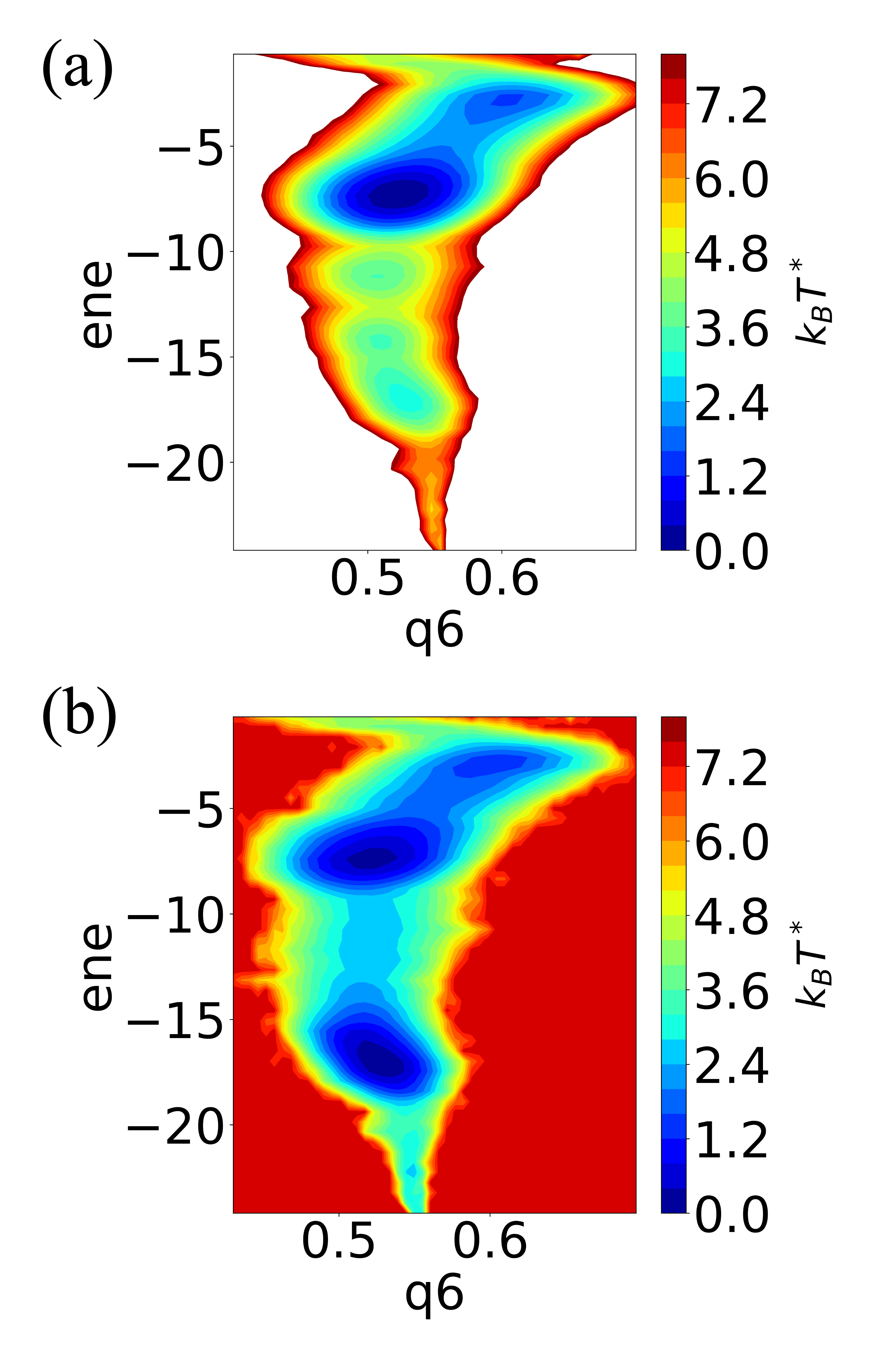}
    \caption{Free energy projection onto q6 and energy for (a) 1000 and (b) 21 unbiased simulations.}
    \label{fig:fe_unbiased}
\end{figure}

In Fig. \ref{fig:fe_unbiased}(a) we provide the free energy approximated from all the initial unbiased simulations, irrespective of whether they crystallized or not, projecting onto the energy and q6 OPs. 
Also shown in Fig. \ref{fig:fe_unbiased}(a) are three metastable crystal basins likewise reported in Ref. \cite{matteo2023}. At $E^*  < -20\epsilon$, we observe the dense crystal structure (Fig. \ref{fig:phases}(d)) in three instances of the crystallized trajectories. However, given enough simulation time, the system inevitably reaches this phase as shown by the unbiased trajectory in Fig. \ref{fig:cn_metad}. The transition from crystal to dense crystal ($E^* \sim  -20\epsilon$) also accounts for a high energy barrier in the free energy profile.  

In Fig. \ref{fig:fe_unbiased}(b) we provide the same approximate free energy obtained by considering only the 21 trajectories that displayed crystallization. These are the trajectories used as input to SPIB. Compared to Fig. \ref{fig:fe_unbiased}(a), there is a singular crystal basin which has similar depth to the liquid basin and the barrier between them is relatively high although not as pronounced as in Fig. \ref{fig:fe_unbiased}(a). Interestingly, both Figs. \ref{fig:fe_unbiased}(a) and \ref{fig:fe_unbiased}(b) have a similar barrier height for the crystal-to-dense crystal transition while both vapor-to-liquid and liquid-to-crystal barriers are lower when only considering crystallized trajectories. This is quite likely due to the poor sampling of this region as the likelihood of the system exploring this space within the simulation time is more improbable than the system crystallizing ($\sim 2 \%$ vs $0.3\%$). This point is highlighted in our metadynamics results. We emphasize that Fig. \ref{fig:fe_unbiased}(a) and (b) are just approximations to the free energy for the sake of visualization and developing intuition, as these simulations display hysteresis and do not correspond to equilibrium. We provide the full converged free energy in the next sub-section, obtained using SPIB-biased metadynamics.

\subsubsection{Metadynamics biasing SPIB}

\begin{figure}[h!]
    \centering
    \includegraphics[width=0.9\linewidth]{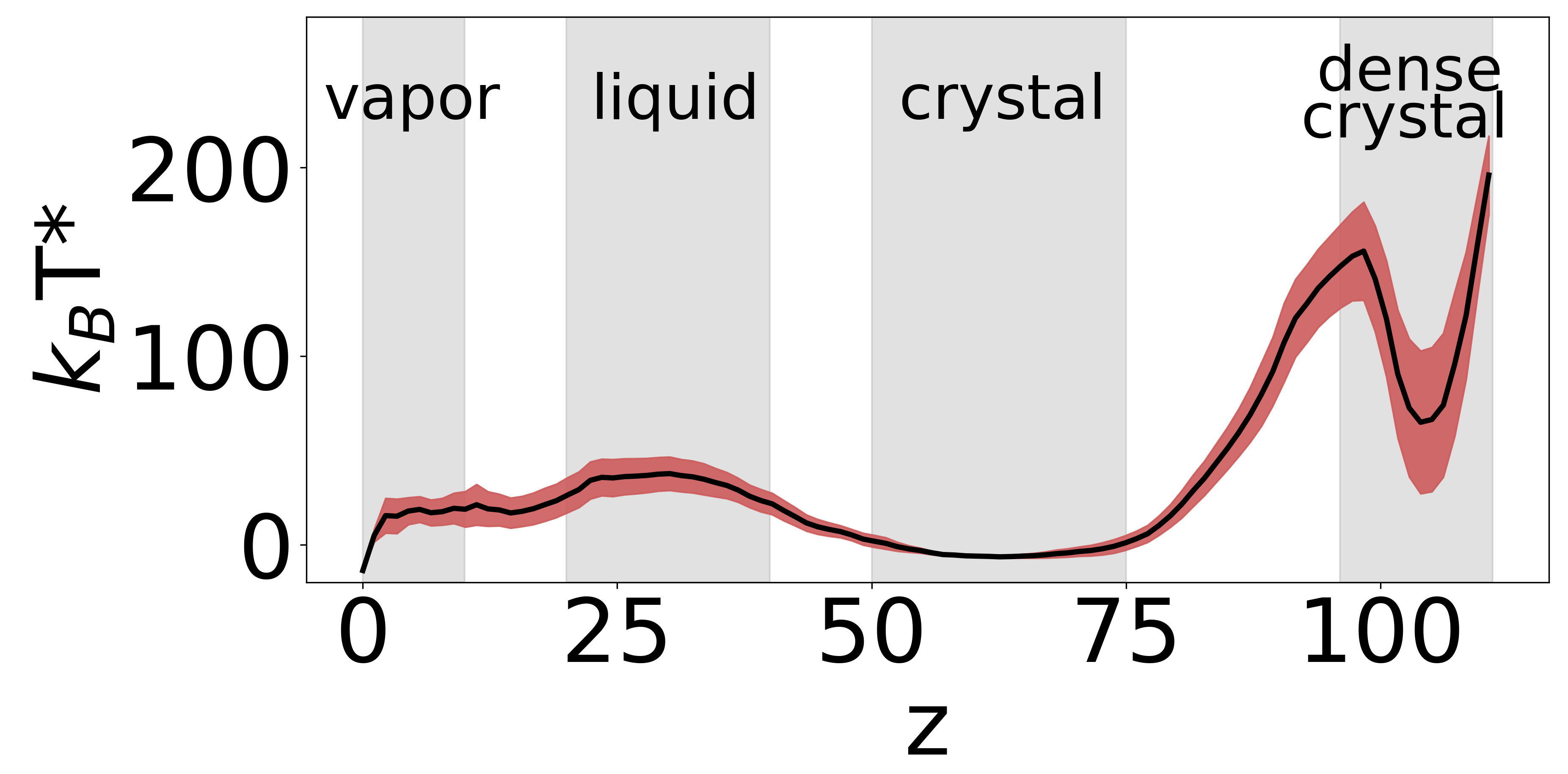}
    \caption{One-dimensional free energy profile from metadynamics in reduced units as a function of the SPIB coordinate $z$.}
    \label{fig:fe}
\end{figure}

We calculate a one dimensional free energy profile for the colloidal system from each of the 10 metadynamics trajectories by reweighting \cite{Tiwary2014}.
Fig. \ref{fig:fe} provides the one-dimensional averaged free energy (in black), the error bars between them (in red), and the classification of phases as determined by OPs. The boundaries of the phases themselves are loosely based on the one-dimensional free energy projection of the OPs, provided in more detail in SI. Understandably, there is a correlation between the OP weights in Fig. \ref{fig: RC weights} and the free energy projections which allow us to reliably define the phase along SPIB coordinate $z$. 

\begin{figure}[h!]
    \centering
    \includegraphics[width=.8\linewidth]{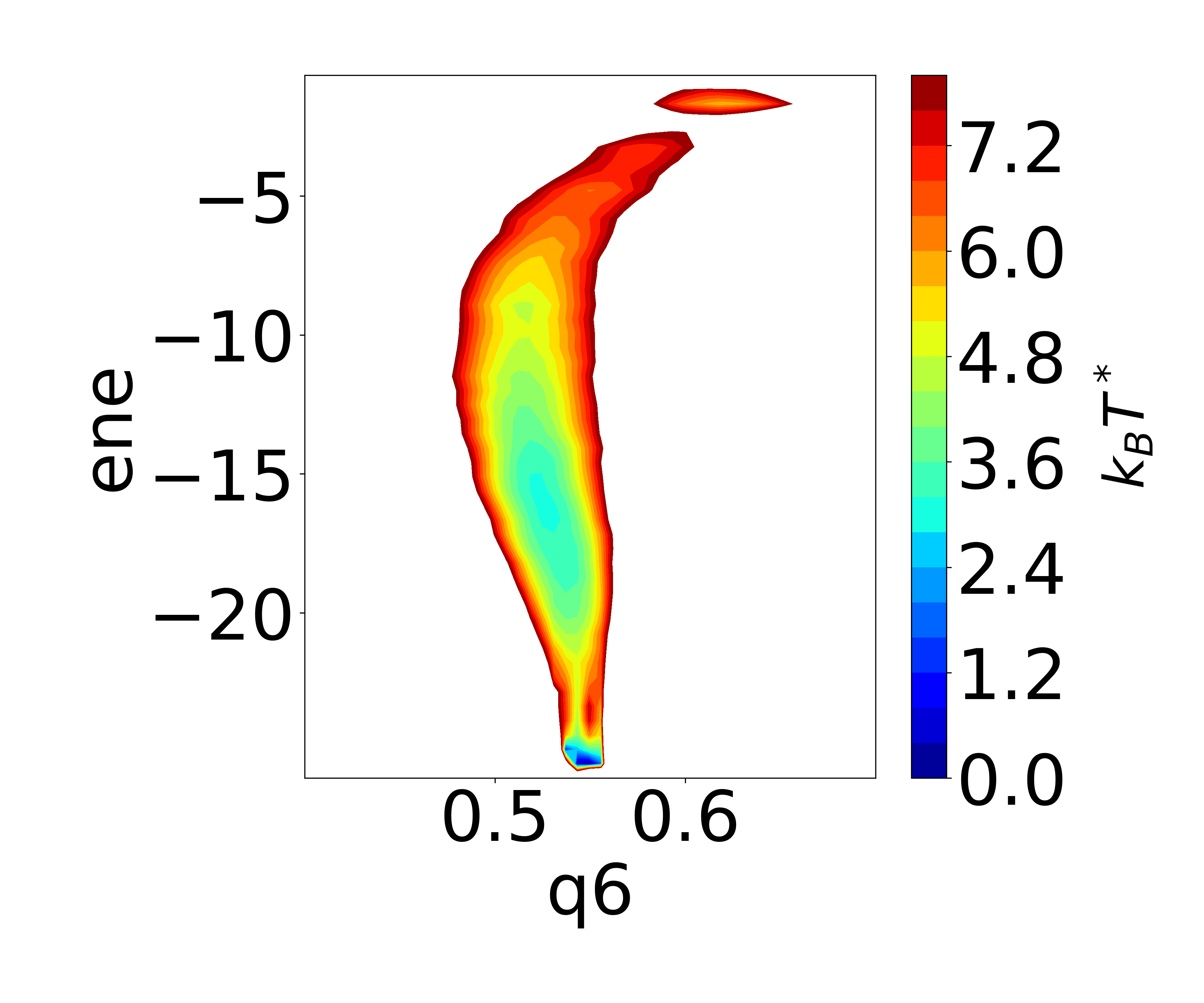}
    \caption{Free energy projection onto Stenihardt bond order parameter q6 and potential energy from metadynamics simulations.}
    \label{fig:fe_metad}
\end{figure}

In  Fig. \ref{fig:fe_metad} we provide a two-dimensional projection of the free energy along coordinates q6, Steinhardt bond order parameter, and ene, average potential energy. This shows significant differences relative to the initial unbiased projection in Fig. \ref{fig:fe_unbiased}. Remarkably, there are two main wells: that of the crystal and dense crystal phase. Just above in the island where $E^* > -5 \epsilon$, the vapor phase is seen. The barrier height difference between the crystal and dense crystal phase in Fig. \ref{fig:fe_metad} is much more pronounced in Fig. {\ref{fig:fe}}. This also explains why once the system crosses into the dense crystal phase it is markedly more difficult for it to go back to the crystal phase.  

\section{Conclusion}
\label{sec:conclusion}

In this article, we have used machine learning augmented molecular dynamics simulations to study the emergence of different ordered phases in a system of supersaturated colloid suspensions. Specifically, we use the State Predictive Information Bottleneck (SPIB)\cite{Wang2021} for learning a one-dimensional reaction coordinate (RC). We see that such a simplified RC is effective in capturing phase transitions in well-tempered metadynamics simulations. This allows us to observe various stable states relevant to the crystallization process in colloidal systems which are identified as rare events under conventional molecular dynamics simulations. Generally, our simulations display multiple back-and-forth transitions between the vapor, liquid and crystal phases.  From the vapor state, we see the system initially shifting to a dense liquid droplet stage before entering a crystalline phase. In addition, there are seemingly direct to-and-from vapor-crystal transitions with only a brief, if any, time spent in the dense liquid phase. Furthermore, it's common for the system to progress into an even denser crystal state where it remains after it's been reached.
This reflects both the strengths and the limitations of our approach. 
We believe the methodology demonstrated in this study will prove valuable in gaining thermodynamic insights into phase transitions in various colloidal systems with the introduction of linear or non-linear combinations of conventional order parameters. To address a broader interest in the study of nucleation problems which, in general, require a larger simulation cell, our protocol can be readily applicable.

%

\section*{Acknowledgments}
\label{sec:acknowledgements}
This work was entirely supported by the U.S. Department of Energy under Award Number DE-SC0021009. We thank UMD HPC’s Zaratan and NSF ACCESS (project CHE180027P) for computational resources. Pratyush Tiwary is an investigator at the University of Maryland-Institute for Health Computing, which is supported by funding from Montgomery County, Maryland and The University of Maryland Strategic Partnership: MPowering the State, a formal collaboration between the University of Maryland, College Park and the University of Maryland, Baltimore.

\bibliographystyle{achemso}
\bibliography{references}

\end{document}




\title{\LARGE Supplementary Information to ``Simulating Crystallization in a Colloidal System Using State Predictive Information Bottleneck based Enhanced Sampling"}

\author{Vanessa J. Meraz}
\affiliation{Institute for Physical Science and Technology, University of Maryland, College Park 20742, USA.}

\author{Ziyue Zou}
\affiliation{Department of Chemistry and Biochemistry, University of Maryland, College Park 20742, USA.}

\author{Pratyush Tiwary\thanks{ptiwary@umd.edu}}
\email{ptiwary@umd.edu}
\affiliation{Institute for Physical Science and Technology, University of Maryland, College Park 20742, USA.}
\affiliation{Department of Chemistry and Biochemistry, University of Maryland, College Park 20742, USA.}
\affiliation{University of Maryland Institute for Health Computing, Rockville 20852, USA.}

\maketitle
\thispagestyle{fancy}



\section{Metadynamics Trajectories}

\label{sec:metad_traj}
\begin{figure}[h!]
\centering
\includegraphics[width=0.44\linewidth]{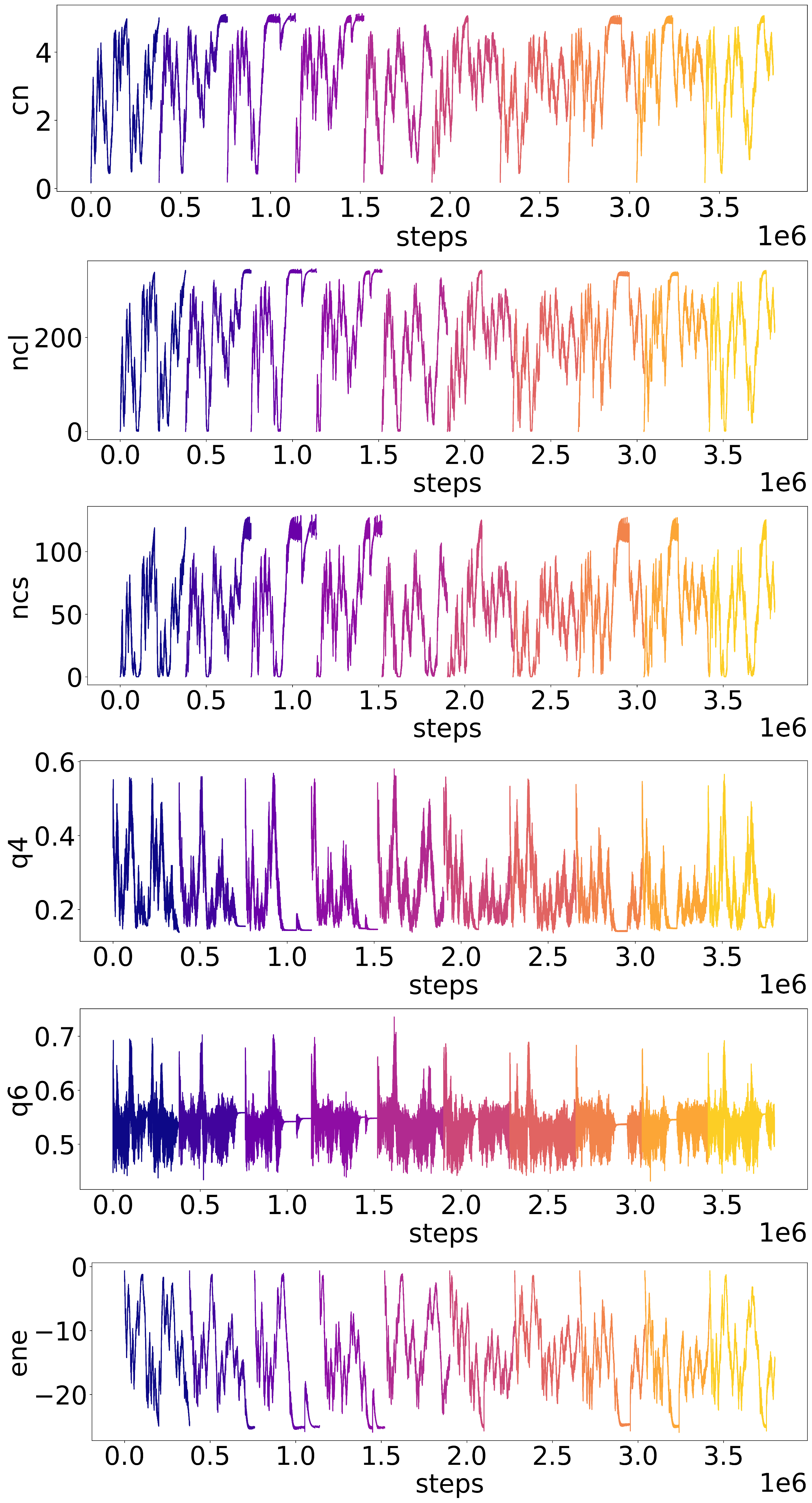}

\caption{Each row shows an order parameter plot for 10 concatenated metadynamics trajectories discretized by color. As mentioned in the main text, there is clear back-and-forth movement between phases seen clearly several OPs. 
Within their respective plots, we can note the dense crystal phase formation briefly in a couple of trajectories where cn $>$ 4 or ene $< -20 \epsilon$. }
\label{fig:metad}
\end{figure}

\newpage
\section{Order Parameter Histograms} 
\label{sec:hist_op}

\begin{figure}[h!]
\centering
\includegraphics[width=0.6\linewidth]{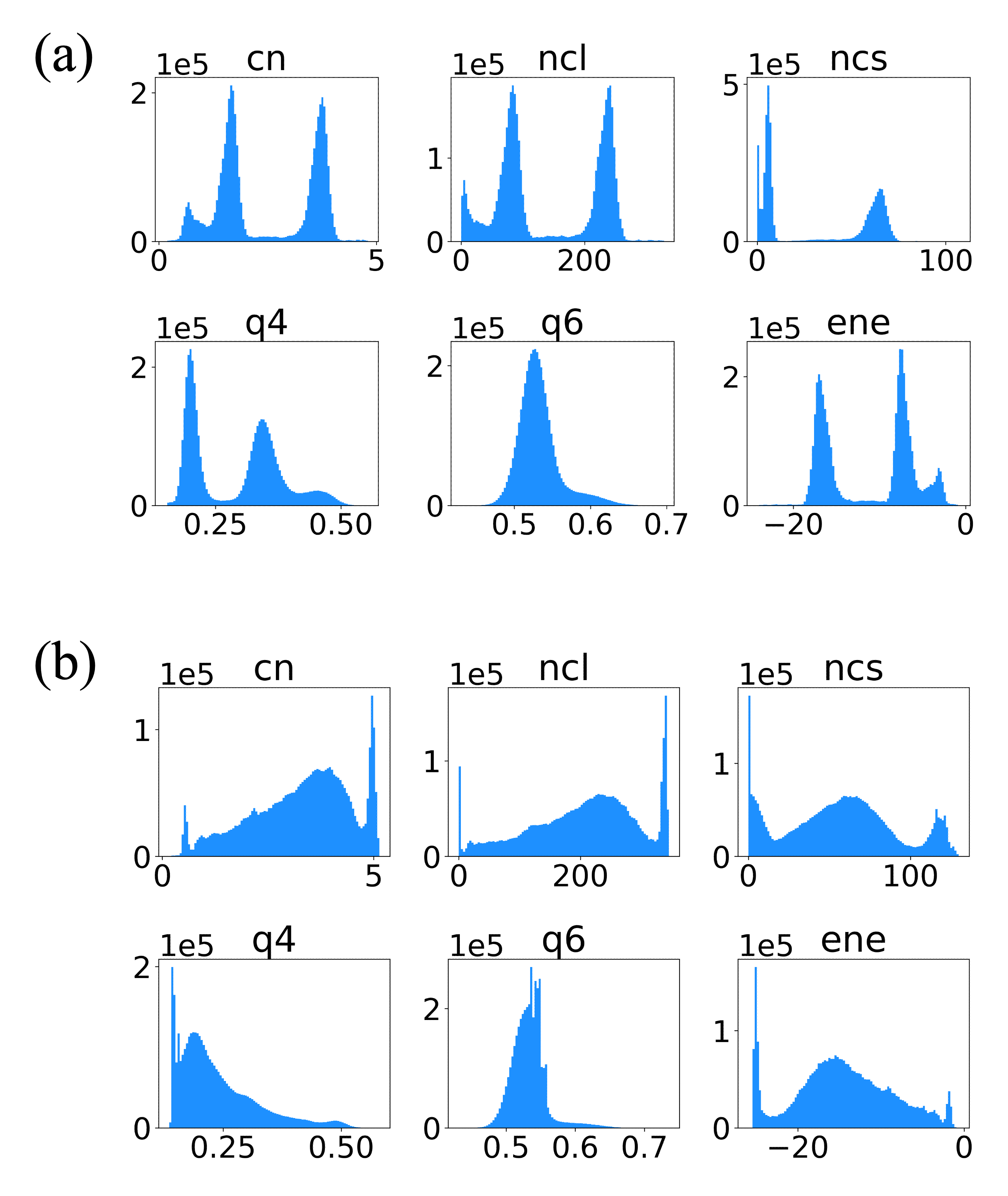}
\caption{In this figure we provide for the sake of completeness direct histograms for different OPs (without any correction for biasing) from (a) unbiased simulations and (b) biased metadynamics runs.}
\label{fig:hists}
\end{figure}

\newpage
\section{One-dimensional free energy profile for all Order Parameters}
\label{sec:1d_fep}

\begin{figure}[h!]
\centering
\includegraphics[width=0.9\linewidth]{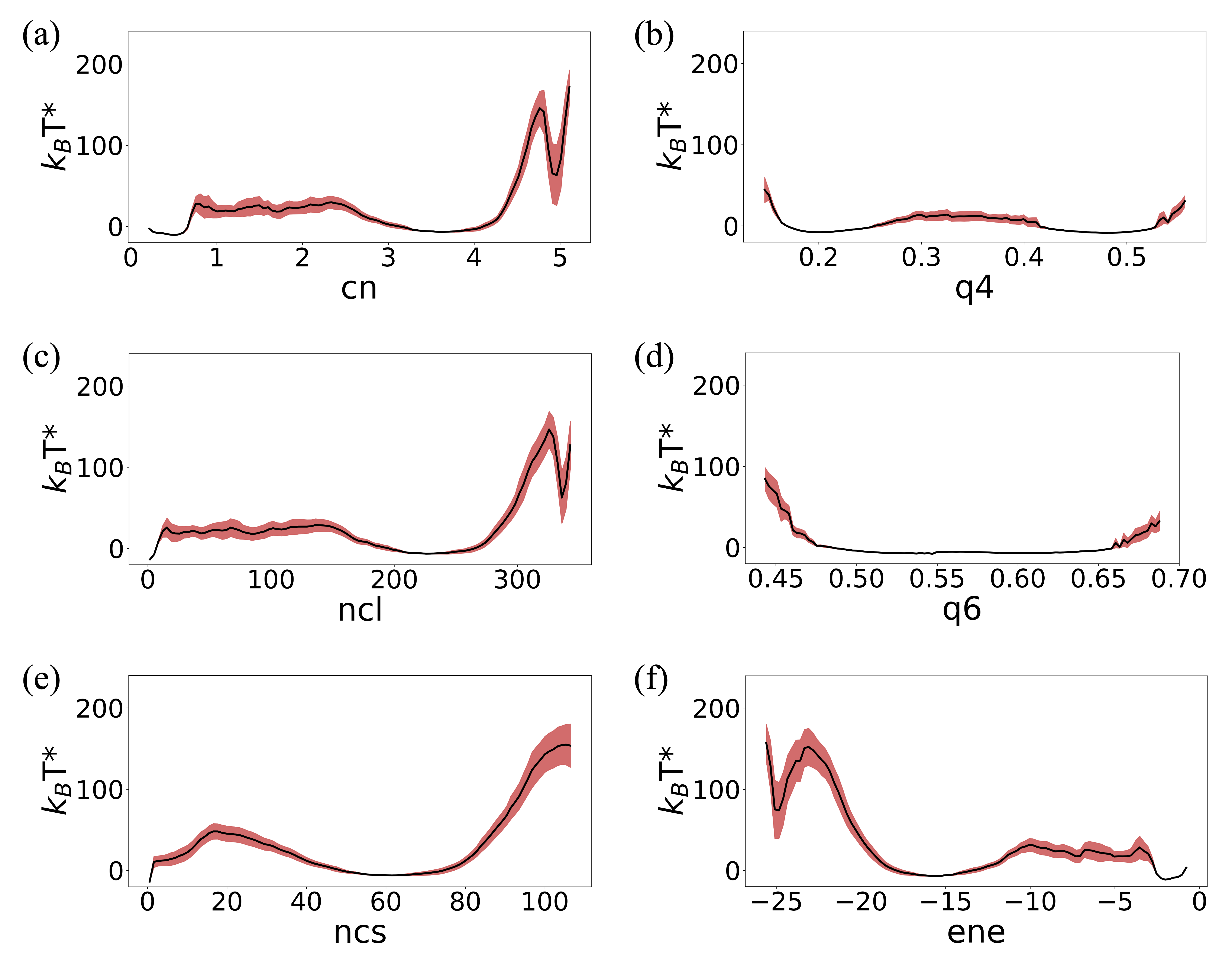}
\caption{In this figure we provide one-dimensional free energy profile in units of $k_BT^{*}$ for each OP. In agreement with Fig. 4 from the main text, which shows the weights of the OPs from the learned SPIB RC, we can observe similarities in the structure of the SPIB $z$ coordinate free energy profile shown in Fig. 8 and the individual ones plotted here. We highlight that the two highest contributors to the RC, the mean coordination number (cn) and the number of particles in a solid-like phase (ncs), have similar features in their free energy profile to that of $z$. Notably, (e) is missing the well characteristic of the dense crystal phase which is otherwise seen in (a), (c), and (f). }
\label{fig:1d_fep_ops}
\end{figure}

\newpage
\section{Pair-plots for two-dimensional free energy projections} 
\label{sec:pp}

\begin{figure}[h!]
\centering
\includegraphics[width=0.9\linewidth]{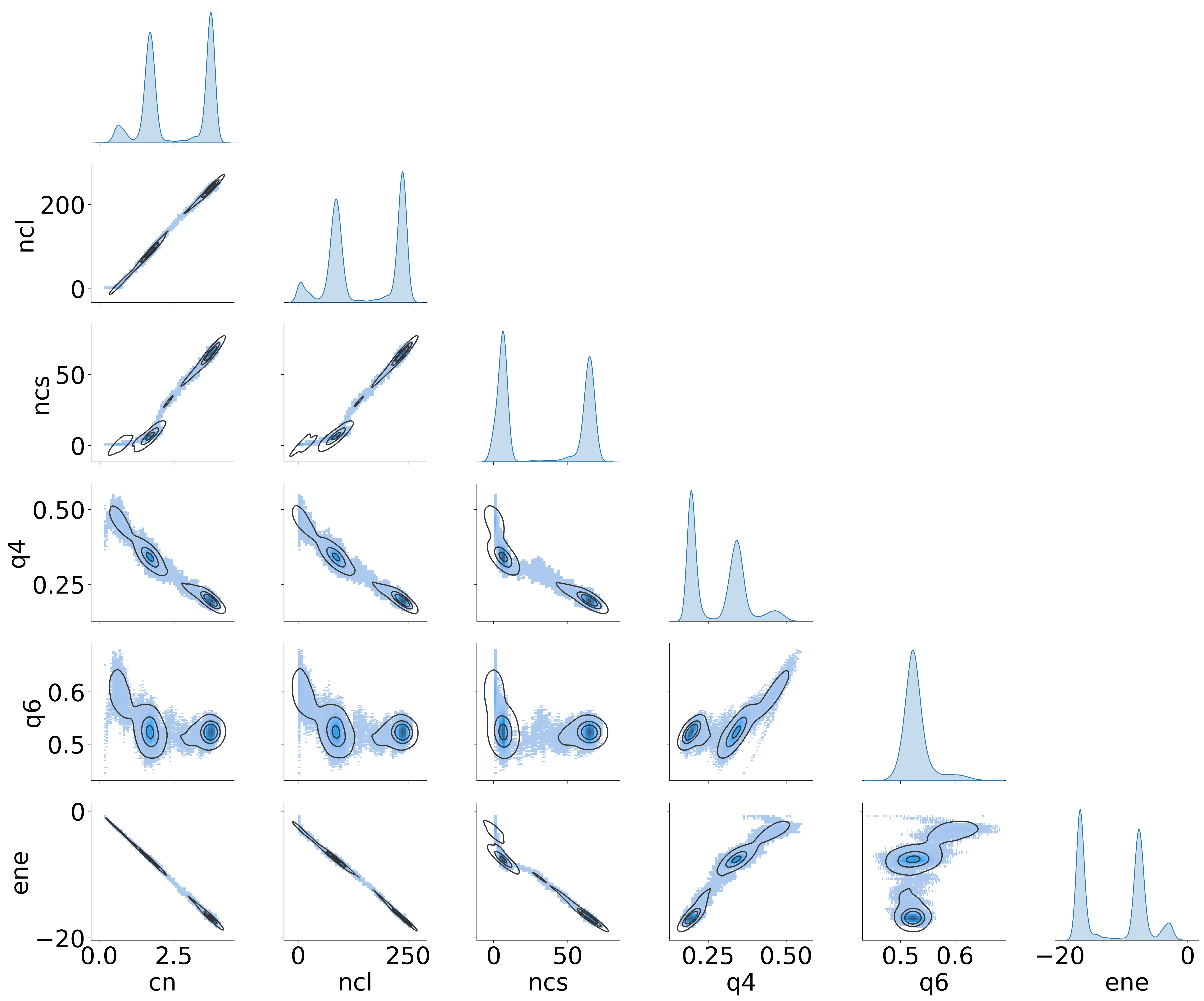}

\caption{As an aid to pick the best representation of a two-dimensional free energy landscape, the pair-plot generated by seaborn in Fig. \ref{fig:pair_plot} was used. Seemingly, quite a few order parameters are highly correlated which don't allow for an easy insight into the systems' phase. Though not as easy to interpret, Steinhardt bond order parameter q6 is the most expressive OP, in which three wells to represent the phases of the system can be seen. Hence, the projection onto q6 and energy is most useful to us and is provided in the main text. Across the diagonal of Fig. \ref{fig:pair_plot} are kernel density estimations (KDEs) for the respective column-wise OP. These are also analogous to the histograms shown in Fig. \ref{fig:hists}(a). Overlaid on the blue tinted histogram plots in the lower triangle shown in black lines are KDEs.}
\label{fig:pair_plot}
\end{figure}


\FloatBarrier
